\PassOptionsToPackage{dvipsnames}{xcolor}
\PassOptionsToPackage{footnote}{nicematrix}
\documentclass[sigconf, review=false]{acmart}
\renewcommand\footnotetextcopyrightpermission[1]{} % removes footnote with conference information in first column

\usepackage{tabu}
\usepackage{xspace}
\usepackage{multirow}
\usepackage{multicol}
\usepackage[font=small]{caption} % decreasing the font size of the caption of table or figure
\usepackage{float}
\usepackage{csquotes}
\usepackage{textcomp}

\usepackage{lipsum}
\usepackage{mwe}
\usepackage{url}
\usepackage[boxed,vlined]{algorithm2e}
\usepackage{paralist} % also compact lists
\usepackage{listings} % to get sql etc nicely formatted
\usepackage{bbding} % for the \CheckmarkBold symbol
\usepackage{hyperref}
\usepackage{array}
\newcolumntype{P}[1]{>{\centering\arraybackslash}p{#1}}

\usepackage{diagbox}
\DeclareUnicodeCharacter{2212}{-}

\usepackage{nicematrix} % for table
\usepackage{tablefootnote} % for table
\usepackage{graphicx} 
\usepackage{subcaption} % for showing figure side by side

\usepackage{tikz}
\usetikzlibrary{matrix}
\usepackage[T1]{fontenc} % Enable custom fonts

\usepackage{mdframed}
\usepackage{xcolor}
\usepackage{roboto} % Load the Roboto font package

\definecolor{lightgray}{RGB}{240, 240, 240}
\definecolor{darkgray}{RGB}{120, 120, 120}

\newenvironment{chatbotresponse}{%
  \begin{mdframed}[backgroundcolor=lightgray,linewidth=0pt]%
}{%
  \end{mdframed}%
}

\newcommand{\hide}[1]{}

\newcommand{\bit}{\begin{compactitem}}
\newcommand{\eit}{\end{compactitem}}
\newcommand{\ben}{\begin{compactenum}}
\newcommand{\een}{\end{compactenum}}

\newcommand{\totalrepo}{{35.2K}\xspace}
\newcommand{\exprepo}{{22.2K}\xspace}
\newcommand{\malrepo}{{9294}\xspace}
\newcommand{\malrepoper}{{26}}

\newcommand{\categorycount}{{14}\xspace}
\newcommand{\keyloggercount}{{1071}\xspace}
\newcommand{\maledu}{{MalEdu}\xspace}
\newcommand{\samplecount}{{100}\xspace}
\newcommand{\accuracy}{{85}}

\newcommand{\PreserveBackslash}[1]{\let\temp=\\#1\let\\=\temp}
\newcolumntype{C}[1]{>{\PreserveBackslash\centering}p{#1}}
\newcolumntype{R}[1]{>{\PreserveBackslash\raggedleft}p{#1}}
\newcolumntype{L}[1]{>{\PreserveBackslash\raggedright}p{#1}}
{}% for the vertical padding
\definecolor{tableHeaderColor}{RGB}{102,204,0}
\definecolor{tableSubHeaderColor}{RGB}{230, 220, 102}
\definecolor{cellCategoryColor}{RGB}{255, 220, 102}
\definecolor{lightgoldenrodyellow}{rgb}{0.98, 0.98, 0.82}
\definecolor{mascommentcolor}{rgb}{0.66, 0.13, 0.24}
\definecolor{masokcolor}{rgb}{0.45, 0.31, 0.59}
\newcommand{\miii}[1]{{\color{purple}{\bf MF:}}
{\color{blue}{\bf #1}}}

\newcommand{\miok}[1]{ {\color{ForestGreen}{{\color{purple}{\bf MF:}} #1}}}

\newcommand{\mas}[1]{{\color{mascommentcolor}{\bf MA:}}
{\color{mascommentcolor}{\bf #1}}}

\newcommand{\masok}[1]{{\color{masokcolor}{\bf MA:}}
{\color{masokcolor}{ #1}}}

\newcommand{\bt}[1]{{\color{RoyalBlue}{\bf BT:}} % Ben
{\color{red}{\bf #1}}}
\newcommand{\btok}[1]{{\color{RoyalBlue}{\bf BT:}}
{\color{ForestGreen}{ #1}}}

\newcommand{\eatreminders}[1]{
\renewcommand{\miii}[1]{}
\renewcommand{\miok}[1]{#1}
\renewcommand{\mas}[1]{}
\renewcommand{\masok}[1]{#1}
\renewcommand{\bt}[1]{}
\renewcommand{\btok}[1]{#1}
}
\eatreminders{}

\begin{document}

\title{
{Unveiling A Hidden Risk: Exposing Educational but Malicious Repositories in GitHub
}}

\author{Md Rayhanul Masud}
\affiliation{%
  \institution{UC Riverside}
    \country{}  
}
\email{mmasu012@ucr.edu}

% \author{Ben Treves}
% \affiliation{%
%   \institution{UC Riverside}
%   \country{}
%   }
% \email{btrev003@ucr.edu}

\author{Michalis Faloutsos}
\affiliation{%
  \institution{UC Riverside}
    \country{}  
}
\email{michalis@cs.ucr.edu}

\thispagestyle{plain}

\begin{abstract}

%%%%%%%%%%%%%%%%%%%%%%%%%%%%%%%%%%%%%%%
% AUTHOR: Christos Faloutsos
% INSTITUTION: CMU
% DATE: April 2019
% GOAL: to streamline the paper presentations
%%%%%%%%%%%%%%%%%%%%%%%%%%%%%%%%%%%%%%%
% \comment{test}
%\notice{{\em rhetorical question:} - What is the best rhetorical question you can start with?}
% \begin{abstract}
% How can we identify similar repositories and clusters among a large online archive, such as GitHub? 
% Can GitHub repositories that are self-promoted as ``for educational purpose only" be considered inherently harmless? This is the question that motivates our work.
Are malicious repositories hiding under the educational label in GitHub? 
% Many online archives, such as GitHub, encourage  developers to  copy, reuse, and improve the existing projects.
% Determining repository similarity is an essential building block in studying the dynamics and the evolution of such software ecosystems.
Recent studies have identified collections of GitHub repositories hosting malware source code with notable collaboration among the developers. Thus, analyzing GitHub repositories deserves inevitable attention due to its open-source nature providing easy access to malicious software code and artifacts.  
%  task for developers where Machine Learning (ML) approaches can play a vital role. 
% The key challenge is to determine the right representation for the diverse repository  features in a way that: (a) it captures all aspects of  the available  information, and (b) it is readily usable by ML algorithms.
%numeric features maintaining the semantics of the content of a repository. %  However, to use standard ML approaches, we need to represent a repository in numeric features maintaining the semantics of the content of a repository. 
% %\notice{{\em 'what' - NOT 'how':} 
%List the benefits of the approach - NOT the details of how you do it!}
% We propose \method, a comprehensive embedding approach to represent a repository as a distributed vector by combining features from three  types of information sources.
Here we leverage the capabilities of ChatGPT in a qualitative study to annotate an educational GitHub repository 
% as malicious/benign/gray-area. 
based on maliciousness of its metadata contents.
Our contribution is twofold. First, we demonstrate the employment of ChatGPT to understand and annotate the content published in software repositories. Second, we provide evidence of hidden risk in educational repositories contributing to the opportunities of potential threats and malicious intents.
We carry out a systematic study on a collection of \totalrepo  GitHub repositories claimed to be created for educational purposes only. First, our study finds an increasing trend in the number of such repositories published every year. Second, \malrepo of them are labeled by ChatGPT as malicious, and further categorization of the malicious ones detects \categorycount different malware families including DDoS, keylogger, ransomware and so on. Overall, this exploratory study flags a wake-up call for the community for better understanding and analysis of software platforms. 
\vspace{-0.8cm}

\end{abstract}

\maketitle

%\section{Acknowledgments}
%We would like to thank blah blah

% \begin{IEEEkeywords}
% Embedding, GitHub, Similarity, Clustering, Software.
% \end{IEEEkeywords}

\section{Problem Definition}
\label{sec:problemdef}
Are GitHub repositories enabling the spread of malware? This is the question that motivates our work. GitHub is the most widely used open source software platform. There are more than 28M public repositories in GitHub ~\cite{githubwiki}, among them 7.5K repositories are identified to contain malware source code; according to a recent study~\cite{rokon2020sourcefinder}. It clearly indicates that public GitHub repositories can host malicious contents. 
As a result, malicious repos can be published in the following ways; (a) repositories are self-determined to be educational publishing proof-of-concept of vulnerabilities and exploits~\cite{githubpolicy}, and (b) sometimes they can intentionally contain malwares ~\cite{fakereseachernews}. Another way is to do that, they may share malicious contents, but promoting as "for educational purpose only". We illustrate such an example of an educational GitHub repository in Figure \ref{fig:work_flow}(a) that shares source code of a ransomware application; however the educational intent does not prevent any malicious actor from using it in an unwanted manner. We refer to these repositories as {\bf \maledu} for the rest of the paper. \\
% Are malware repositories hiding under the educational label?
% How can we find educational, but malicious repositories in GitHub? 
The problem we address here is the following: given a GitHub repository that is self-promoted as published for educational purpose only, how can we determine whether it is likely to be malicious? So,
the input to the problem is GitHub, and the expected output is a set of \maledu repos. The challenges include: (a) collecting educational repos, and (b) identifying the malicious ones among them.

\begin{figure}[t]
    \begin{center}
        \begin{tabular}{c}
          \includegraphics[width=0.42\textwidth]{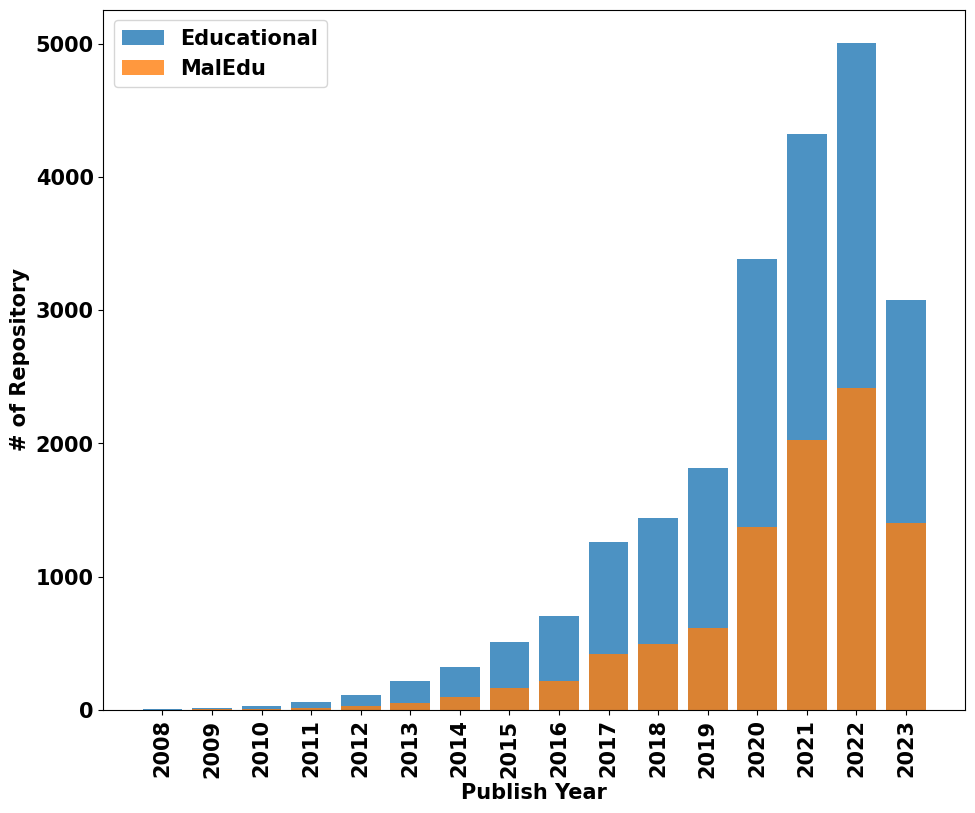} \\
        \end{tabular}
        \captionsetup{skip=0pt} % Set the skip parameter to adjust the spacing
        \caption{The number of educational GitHub repositories is increasing every year. The trend is similar for \maledu (educational, but malicious) repositories. 
          \label{fig:temporal} }
    \end{center}
\vspace{-4mm}
\end{figure}

\section{Contribution}
\label{sec:contribution}

%%%%%%%%%%%%%%%%%%%%%%%%%%%%%%%%%%%%%%%
% AUTHOR: Christos Faloutsos
% INSTITUTION: CMU
% DATE: April 2019
% GOAL: to streamline the paper presentations
%%%%%%%%%%%%%%%%%%%%%%%%%%%%%%%%%%%%%%%

As our key contribution, we propose a systematic study to analyze educational GitHub repositories to identify the repositories that contain malicious contents. We apply our method on a collection of \totalrepo educational GitHub repositories (excluding forks) published during the period between 2008 and 2023 (Jun 24). Our key results are briefly discussed below.

% we identify \malrepo educational repositories labeled and categorized into \categorycount malware families with the help of ChatGPT. 

{\bf a.} The number of educational repositories in GitHub has been increasing each year since its launch. According to figure \ref{fig:temporal}, the frequency of the repositories published during 2020 and 2023 is 2.4 times the total number of repositories published before.

{\bf b.} We find \totalrepo educational repositories. \malrepo ($\sim$\malrepoper\%) of them are identified as {\maledu} repositories. 
% based on unanimous decision in a two-phase query execution pipeline powered by ChatGPT, which we describe in detail in Section \ref{sec:exp}.
Further categorization of \maledu repos finds \categorycount different malware families.

{\bf c.} Our manual validation suggests that ChatGPT accurately detects \maledu repositories with \accuracy\% precision.

\section{Methodology}
\label{sec:exp}

% %%%%%%%%%%%%%%%%%%%%%%%%%%%%%%%%%%%%%%%
% % AUTHOR: Christos Faloutsos
% % INSTITUTION: CMU
% % DATE: April 2019
% % GOAL: to streamline the paper presentations
% %%%%%%%%%%%%%%%%%%%%%%%%%%%%%%%%%%%%%%%

% % \notice{{\em novelty}: NO citations, outside the 'survey'  - 
% % they make \method seem incremental. }

% \subsection{Dataset}
{\bf A. Data collection.} We use the GitHub search API to collect educational repositories. A GitHub repository has multiple metadata fields including repo title, description, readme file, star/fork/watch count and so on. We query the search API for repositories (excluding forks) with the following phrases in description and readme content; (a) education/educational purpose only, (b) only for 
education/educational purpose. This yields a collection of \totalrepo educational repositories. Then, we filter the repos that contain both description and readme content, which results in \exprepo repositories that we consider in our experiment. \\
\begin{figure}[t]
        \centering
        \subfloat[Example of a \maledu GitHub Repository.]{{\includegraphics[width=0.6\columnwidth]{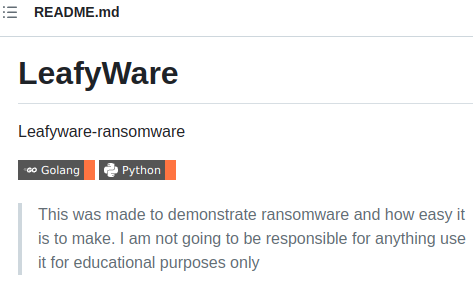}}}
        \vspace{2mm}
        
        \subfloat[Visualization of our approach.]{{\includegraphics[width=0.95\columnwidth]{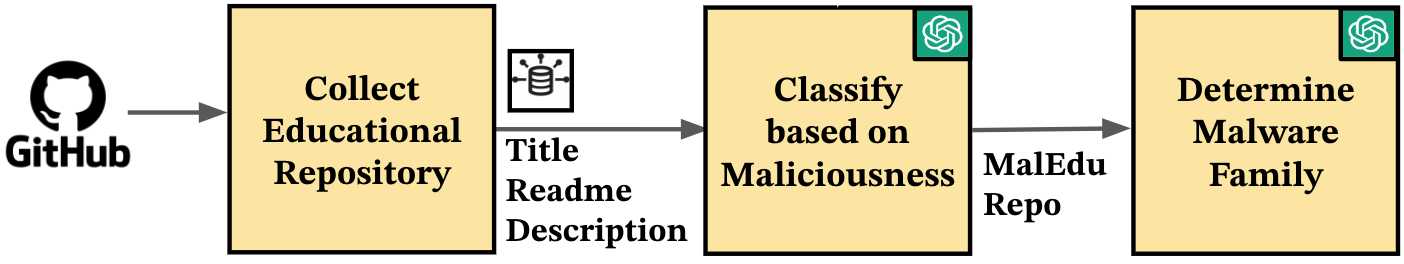} }}

        \captionsetup{skip=2pt} % Set the skip parameter to adjust the spacing
        \caption{(a) Example metadata of GitHub repository that hosts ransomware source code, while created for educational purpose only. (b) First, a collection of educational repos is classified by ChatGPT. Then, identified \maledu repos are classified into malware families.}
        
        \label{fig:work_flow}
    \vspace{-8mm}
\end{figure}
% \subsection{Workflow}
{\bf B. Determine maliciousness.} Our aim is to determine how many educational repositories in GitHub platform can be labeled as malicious. A recent study ~\cite{zuccon2023dr} has found ChatGPT quite effective for answering health related questions (Yes/No) relying solely on the model knowledge. It motivates us to engage ChatGPT to classify the contents of a repository to identify the \maledu repositories.

% \begin{figure}[t]
%     \begin{center}
%         \begin{tabular}{c}
%           \includegraphics[width=0.47\textwidth]{FIG-POSTER/work_flow.png} \\
%         \end{tabular}
%         \caption{Visualization of our method. Filtered educational repositories are classified into three categories (We query ChatGPT twice). ChatGPT labeled malicious repos are further categorized into granular types.
%           \label{fig:work_flow} }
%     \end{center}
% \vspace{-6mm}
% \end{figure}
\vspace{2ex}
\begin{chatbotresponse}
\underline{\bf ChatGPT Prompt:} \\
\textbf{\textit{Context:}} Say you are a security professional. 
Given specific information about a repository, such as repo title, description and the readme file content, 
you will annotate the repository whether the repo is malicious. \\
\textbf{\textit{User:}} \\
Repository Title: ...\\ 
Description: ...\\
Readme File Content: ... \\
Based on the provided information, please annotate with one option: benign, malicious, gray-area;  
indicating the potential maliciousness of the repository. No explanation needed. \\
\textbf{\textit{ChatGPT:}} benign/malicious/gray-area.
\end{chatbotresponse}
\vspace{2ex}
We use ChatGPT API based on gpt-3.5-turbo-0613 to annotate the repositories in our dataset. We ask ChatGPT to choose one label among three categories; (a) benign, (b) malicious, and (c) gray-area; given title, readme content and description of a given repository indicating the maliciousness. Since ChatGPT is a generative model, we run two independent queries for each repository annotation, as suggested in a recent study ~\cite{cao2023study}. Thus, we obtain two annotations from two queries for each of the repositories in our dataset. We extract the repositories which get a unanimous decision on the category ``malicious" that provides us the list of \maledu repositories. \\
{\bf  C. Determine malware family.} We want to determine the content type of the identified \maledu repositories. To achieve this goal, first we create a list of popular malware families. Then, we ask ChatGPT to choose a family from the list for a given \maledu repository. If the repository cannot be labeled using the list, ChatGPT is instructed to label it as ``Miscellaneous". The detailed workflow is illustrated in Figure ~\ref{fig:work_flow}(b).

% \subsection{Results}
% We plot the number of educational repositories published each year in the span of 16 years (2008-2023) in Figure ~\ref{fig:temporal}. We also plot the distribution for the ChatGPT labeled malicious ones. We find that new repositories are adding every year, however, the increase is quite significant after 2017.

% \begin{figure}[t]
%     \begin{center}
%         \begin{tabular}{c}
%           \includegraphics[width=0.48\textwidth]{FIG-POSTER/temporal_curated.png} \\
%         \end{tabular}
%         \caption{The number of educational GitHub repositories is increasing every year. The trend is similar for the number of ChatGPT labeled educational, but malicious repositories. 
%           \label{fig:temporal} }
%     \end{center}
% \end{figure}

\begin{figure}[t]
    \begin{center}
        \begin{tabular}{c}
          \includegraphics[width=0.3\textwidth]{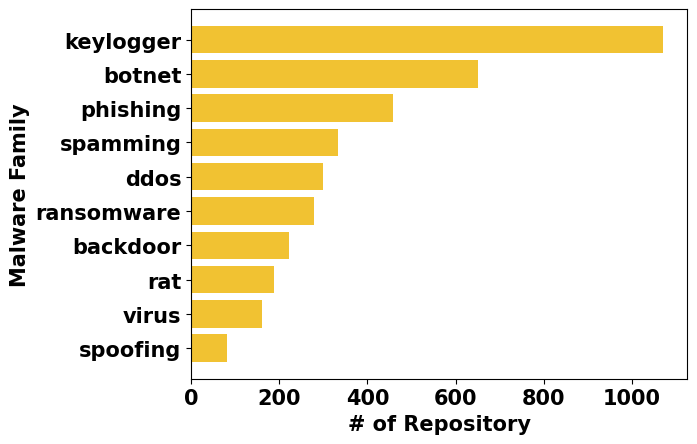} \\
        \end{tabular}
        \captionsetup{skip=0pt} % Set the skip parameter to adjust the spacing
        \caption{Top 10 malware families detected among \maledu repos.
          \label{fig:category} }
    \end{center}
    \vspace{-2mm}
\end{figure}

\section{Results and Evaluation}
\label{sec:results}
\begin{figure}[H]
    \begin{center}
        \begin{tabular}{c}
          \includegraphics[width=0.27\textwidth]{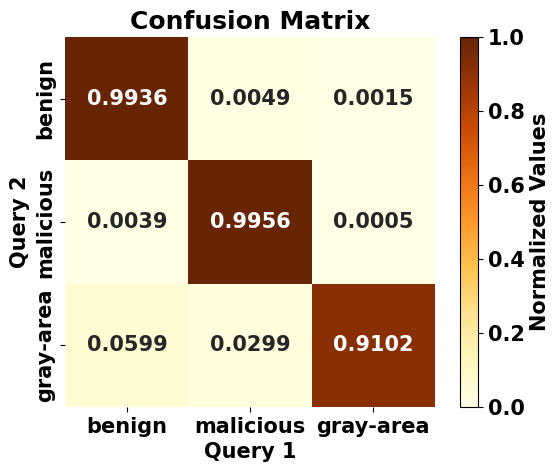}
        \end{tabular}
        \captionsetup{skip=0pt} % Set the skip parameter to adjust the spacing
        \caption{Normalized Confusion Matrix for ChatGPT annotations.
          \label{fig:confusion} }
    \end{center}
    \vspace{-4mm}
\end{figure}

{\bf Results.} We identify \malrepo \ \maledu repositories based on the annotations provided by ChatGPT. The normalized confusion matrix in figure \ref{fig:confusion} shows that
ChatGPT annotations are found to be identical in almost all cases across both query processing. \\
We also detect \categorycount malware families during the categorization of \maledu repository contents. We find ``keylogger" as the most frequently identified malware family accounting to \keyloggercount \ \maledu repositories. Figure \ref{fig:category} lists top 10 malware families detected in this study. \\
{\bf Evaluation.}  To increase our confidence, we randomly select \samplecount ``malicious" labeled (unanimously) GitHub repositories. Then, we investigate the contents of each of them for potential maliciousness. This manual investigation suggests that ChatGPT accurately detects \maledu repositories with \accuracy\% precision.
\vspace{-4mm}

\vspace{-0.2cm}
\section{Future Work}
\label{sec:future}
Following the interesting findings, we intend to take a deep dive into the identified \maledu repos for further profiling their authors and contents. 
We plan to investigate to detect any collaborative approach for the spread of such contents. 
In addition, we also want to verify the functionality of the source code to estimate the potential harm the \maledu repositories can do. 

\vspace{-0.2cm}
\section{Related Work}
\label{sec:related_works}
Though several approaches aim to identify malware repositories in GitHub, none of them considers educational repositories as a possible source of malicious contents.
A recent work, SourceFinder ~\cite{rokon2020sourcefinder} gathers repositories based on keyword search, and then applies machine learning classifier on the repository content embedding to identify malware repos. Another recent study, GitCyber ~\cite{zhang2020cyber} incorporates cybersecurity domain knowledge along with code contents in a deep neural network for malicious repository detection.

\vspace{-0.24cm}
\section{Acknowledgment}
This work was supported by the  NSF SaTC grant No. 2132642.

% \vspace{-0.1cm}
% \section{Conclusion}
% \label{sec:concl}
% \input{060conclusion}

% \input{080listOfToDo}
\newpage
\bibliographystyle{ACM-Reference-Format}
\bibliography{BIB/biblist}

% \reminder{
% \newpage
% \appendix
% \input{070appendix}
% \newpage
% \input{080listOfToDo}
% }

\end{document}